\documentclass[osajnl,twocolumn,showpacs,superscriptaddress,10pt]{revtex4-1} 
\usepackage{amsmath,amssymb,graphicx, verbatim, amsthm, caption, subcaption, float, subfig, mathtools}
\newtheorem{thm}{Theorem}[section]

\newtheorem{cor}[thm]{Corollary}

\begin{document}

\title{Equitable Mirrors}


\author{Casey Douglas}\email{Corresponding author: cjdouglas@smcm.edu}
\affiliation{St. Mary's College of Maryland,  18952 E. Fisher Rd, St. Mary's City, MD 20686, USA}

\author{Emek K\"{o}se}\email{Corresponding author: ekose@smcm.edu}
\affiliation{St. Mary's College of Maryland,  18952 E. Fisher Rd, St. Mary's City, MD 20686, USA}

\author{Nora Stack}\email{Corresponding author: nstack@smcm.edu}
\affiliation{St. Mary's College of Maryland,  18952 E. Fisher Rd, St. Mary's City, MD 20686, USA}

\author{Caroline Van Blargan}\email{Corresponding author: cvanblargen@smcm.edu}
\affiliation{St. Mary's College of Maryland,  18952 E. Fisher Rd, St. Mary's City, MD 20686, USA}

\begin{abstract} In this article we explore some finer properties of equi-areal mirrors and introduce techniques for developing new mirror surfaces that simultaneously minimize angular and areal distortion.

\end{abstract}


\maketitle

\section{Introduction} 

Cartography enjoys a history nearly as old, rich, and problematic as human civilization itself.  Early maps were either small in scope or wildly inaccurate, distorting important properties and lacking in mathematical detail.  The possibility of designing an accurate (i.e. ``distortion free") map of the earth was resolved in the negative by Gauss's {\em Theorema Egregium}:

\begin{thm} (Theorema Egregium) The Gaussian curvature of a surface is invariant under local isometry.
\end{thm} 

Since any function that preserves angles and areas is a local isometry, The Theorem Egregium demands that functions between differently curved surfaces feature area or angle distortion (or both).  This is true for mappings that uniformly rescale area and preserve angles, too.   As a consequence, all differentiable functions between a sphere and a plane either distort angles or areas so that every flat representation of the earth suffers some kind of imperfection.  Cartographers have been left to determine those maps that feature the \emph{least} amount of distortion.

Different maps are commonly chosen based on the needs of their users, then, and this process can be highly subjective.  For instance, navigators preferred the Mercator Projection since angles are displayed accurately \cite{Feeman-book}, whereas equal-area maps are ideal for displaying area-based data.  However, most maps are used exclusively based on their appearance.     
    
While cartographers are interested in mapping the spherical globe to a plane, we are interested in mapping our so-called \emph{object sphere} onto an \emph{image plane}, an essentially equivalent problem.  Our goal is to realize those projections from the sphere to the plane that enjoy a wide field of view but whose \emph{combined} angle and area distortions are minimal. Of the several options available to us, a simple solution is to use a {\em catadioptric sensor}, an optical device that combines a camera with reflective surfaces. 

A camera is a device that maps $\mathbb{R}^3$ to $\mathbb{R}^2$ via a specified projection. 
In our designs we use simplistic models for cameras. The two most commonly used projections are \emph{perspective} and \emph{orthographic}.  A pinhole camera realizes perspective projection, where all light rays pass through a pinhole and form the image. The distance between the pinhole and the image plane is the focal distance. Orthographic projection can be viewed as a limiting case of perspective projection where the focal distance becomes infinite and the light rays travel parallel to the optical axis. 

Much like cartographers' maps, then, cataidoptric sensors are designed based on the needs of their users.  When these needs involve minimizing a single kind of distortion much can be said ~\cite{hb1, hp, chahl_srinivasan, conroy_moore}.  
However minimizing multiple types of distortion has not been as popular. In this paper we propose and analyze precise ways in which to measure and simultaneously minimize multiple kinds of distortion; as an example of this process, we minimize a rotationally symmetric mirror's \emph{total} angular and areal distortion.       

\section{Background} 


The main problem in catadioptric sensor design is to realize a given projection with a camera-mirror pair. In other words, one seeks to determine a mirror surface that, when coupled with a camera, reflects the world in a desired way. 

Mirror surfaces can be designed so that the catadioptric sensor preserves certain geometrical properties. For example, image-to-world mapping can be conformal, equi-areal, equidistant, equiresolution, etc., as in \cite{chahl_srinivasan, conroy_moore, hp, kr}. 
In efforts to achieve uniform resolution, for instance, Conroy and Moore designed catadioptric sensors that have solid angle-pixel density invariance ~\cite{conroy_moore}, while Chahl and Srinivasan study the mirror surfaces that preserve a linear relationship between the angle of incidence of light onto the surface and the angle of reflection ~\cite{chahl_srinivasan}. Neither of these sensors however have a uniformly distributed correspondence between points in the image plane and the object sphere. Hicks and Perline present the family of equi-areal mirrors, so named because they induce projections that uniformly rescale area. ~\cite{hp}.  

The so-called \emph{equitable mirrors} that we seek allow for both area and angle distortion but attempt to minimize their combined, overall impact.  As such, we first review basic and common notions of angle and area distortion.  

\subsection{Angle Distortion}\label{SubSec:Distortion}

Let $K \geq 1$.  Then, as detailed in ~\cite{Ahlfors-book}, an orientation-preserving diffeomorphism $w:U\subseteq\mathbb{C}\rightarrow\mathbb{C}$ is said to be $K$-\emph{quasiconformal} if \begin{equation}\label{quasi}\sup_{z\in U} \frac{\left|w_z\right|+\left|w_{\bar{z}}\right|}{\left|w_z\right|-\left|w_{\bar{z}}\right|}\leq K\end{equation} where we have used the notation \begin{align*} w_z &= \frac{1}{2}\left(w_x - iw_y\right) \,\text{ and } \, w_{\bar{z}} = \frac{1}{2}\left(w_x + iw_y\right). \end{align*}  Additionally, an orientation-reversing diffeomoerphism, $w$, is said to be $K$-quasiconformal if its orientation-preserving conjugate, $\overline{w}$, is $K$-quasiconformal. 

While angle-preserving or conformal mappings take infinitesimal circles to infinitesimal circles, quasiconformal mappings take infinitesimal circles to infinitesimal ellipses.  In fact, the fraction appearing in definition (\ref{quasi}) is precisely the eccentricity of an infinitesimal ellipse centered at the point $w(z)$.  If this eccentricity is one, then $w$ preserves angles at $z$.  If this eccentricity is infinite, then $w$ effectively \emph{destroys} angles at $z$ and fails to be quasiconformal.    

Definition (\ref{quasi}) can be rephrased using the \emph{complex dilatation} of a mapping, $\mu$, which is defined by \begin{equation*}\label{mu}\mu(z) = \frac{w_{\bar{z}}}{w_z} \end{equation*}  A mapping $w$ is $K$-quasiconformal, then, if its complex dilatation satisfies $\|\mu\|_{\infty} = k$ and $$\frac{1+k}{1-k} \leq K.$$  Furthermore, angles are preserved when $\mu=0$ and destroyed when $|\mu|=1$.  

The complex dilatation $\mu$ need only be a measurable, essentially-bounded function, thereby extending the notion of a quasiconformal mapping to a much wider class of functions.  For the purposes of this paper, though, we can restrict the definition to orientation-reversing and orientation-preserving diffeomorphisms.  One readily checks that the dilatation for an orientation-reversing map, $w$, is the reciprocal of the dilatation for $\overline{w}$.  

A rotationally symmetric mirror surface with profile curve $f(r)$ induces a planar mapping between the image plane and the stereographically-projected object sphere.  We use the sphere at infinity to model this object sphere, but because our mirrors are rotationally symmetric, we can replace it with the circle at infinity.   The reversibility of light paths allow us to model the problem by tracing an orthographically-projected light ray.  
The dilatation associated to such a mapping under orthographic projection is given by the formula


\begin{align} 
\label{dilgraph} \mu(r) &= \frac{rf''(r) - f'(r)}{rf''(r)+f'(r)} \end{align}

More generally, if we produce a mirror surface by revolving the planar curve $\gamma(t) = (r(t), z(t))$ about the vertical $z$-axis, then the dilatation of its induced mapping under orthographic projection is given by
\begin{equation}\label{dilcurve} \mu(t) = \frac{(z''\,r'-z'r'')(r'r)-(r')^3z'}{(z''\,r'-z'r'')(r'r)+(r')^3z'}. \end{equation}

For instance, surfaces of the form $f(r) = a r^p$ have constant $\mu = (p-2)/p$ and therefore have $K = p-1$ (for $p\geq2$).  These \emph{Uniform Quasi-Parabaloid} mirrors distort angles by a factor of $p-1$.

\subsection{Area Distortion}\label{SubSec:AreaDistortion}


A catadioptric sensor is called equi-areal if its {\em magnification factor}, denoted by $m_f$, is constant.  This quantity compares the area of a region on the object sphere to the area of its projection on the image plane. Hicks and Perline compute the magnification factor of a sensor under orthographic projection by considering two concentric circles of radii $r$ and $r+\Delta r$ around the optical axis ~\cite{hp}. The infinitesimal annulus with area $\pi((r+\Delta r)^2-r^2)$ is then mapped to the topological annulus on the sphere, bounded by $\phi$ and $\phi+\Delta \phi$, whose area is $\pi[(1-\cos{(\phi+\Delta \phi))})-(1-\cos{(\phi)}).$ The ratio of these two areas, as $\Delta r\to 0$ is the magnification factor

\begin{align*}
m_f =& \displaystyle\lim_{\Delta r\to 0}\frac{\pi(\cos{(\phi)}-\cos{(\phi+\Delta \phi)})}{\pi((r+\Delta r)^2-r^2)}\\
=&\frac{\sin{(\phi)}}{2r}\frac{d\phi}{dr}
\end{align*}

Using the fact that $\phi=\arctan{(f'(r))}$, we obtain the expression for the magnification factor: 

\begin{equation}\label{mf}
m_f = \frac{2f'(r)f''(r)}{r(1+(f'(r))^2)}.
\end{equation}

\section{Preserving Area \emph{or} Angles}\label{Sec:AreaAngle} 

For completeness sake we include a brief but thorough treatment of conformal mirrors.  Although it is well known that the only rotationally symmetric, conformal mirror is the paraboloid, Theorem \ref{Conformal Theorem} classifies \emph{all} conformal mirror surfaces.  


 \begin{thm} \label{Conformal Theorem}The only angle-preserving mirrors (with respect to orthographic projection) are graphs of harmonic functions $z=f(x,y)$ and graphs of the quadric surfaces \begin{equation*} \label{Quadric} z = a\left((x-x_0)^2+ (y-y_0)^2\right) + d \end{equation*} where $a, d \in \mathbb{R}$ and $(x_0,y_0) \in \mathbb{R}^2$ are fixed. \end{thm}  

\proof (sketch) After parameterizing a portion of the mirror surface as a graph, $z=f(x,y)$, we find that the angle-preserving properties are determined by the mapping $g(x,y) = \nabla f(x,y)$.  This follows since stereographic-projection is a conformal mapping between the sphere at infinity and the plane.  In particular, our mirror preserves angles if and only if $$\frac{\partial g}{\partial \overline{z}} = 0 \text{ or } \frac{\partial g}{\partial z} = 0$$ where the first equation implies $g$ is orientation-preserving, and the second one implies it is orientation-reversing.  For the orientation-preversing case, we find
$$ \frac{\partial g}{\partial \overline{z}} = 0 \iff f_{xx} = f_{yy} \text{ and } f_{xy} = -f_{yx} .$$  From these equations it easily follows that $$f(x,y) = a(x-x_0)^2 + a(y-y_0)^2+d$$ where $x_0, y_0, d \in \mathbb{R}$ are fixed. 

Analogous computations confirm that when $g$ is an orientation-reversing, angle-preserving map the function $f(x,y)$ must satisfy \begin{align*} f_{xx} &= -f_{yy} \\ f_{xy} &= f_{yx} \end{align*}  In other words, \emph{$f(x,y)$ is harmonic}. $\square$

Of course, graphs of harmonic functions are not viable as mirror surfaces since they necessarily cause self-reflections.  Moreover, after vertically and horizontally shifting our parabolic mirror, it can be assumed to have the more familiar form $f(r) = ar^2.$  For the remainder of the paper, then, we use $r\in\mathbb[0,\infty)$ as our independent variable and use $y$ to denote the height coordinate of the reflected light ray on the circle at infinity.

The magnification factor for a parabolic mirror is given by $$m_f(r)= \frac{8a^2}{(1+4a^2r^2)^2}.$$  If our parabolic mirror is to have a large field of view over a finite interval $r\in[0,r_1]$, then it will suffer significant area distortion.  This follows from the fact that $$m_f(0) = 8a^2\to\infty\,\text{ as }\,a\to\infty.$$  In order for $f(r)$ to achieve a full field of view at $r=r_1$ the parameter $a$ necessarily tends to infinity.  Moreover, as $m_f(0)$ becomes unbounded, $m_f(r)\to0$ for all other values of $r$.  In summary, then, conformal mirrors with large fields of view have \emph{highly} non-uniform magnification.

At the other end of the spectrum lie the constant-Gaussian curvature surfaces of revolution, mirror surfaces that Hicks and Perline showed are equi-areal (see \cite{hp}).  As categorized in \cite{DiffGeo}, these surfaces fall into one of six families: \emph{spheres, bulges, spindles, psuedospheres, hyperboloid types,} and \emph{conic types}.  The first three have positive Gaussian curvature and are generated by the curve $$\gamma(t)=\left(b\,\cos\left(\frac{t}{a}\right),\int_0^{t/a}\!\sqrt{a^2-b^2\sin\left(\frac{s}{a}\right)}\,ds\right)$$ where $a, b \in (0,\infty)$ are fixed. The latter three are orientation reversing mirrors generated by hyperbolic versions of $\gamma(t)$; these surfaces are similarly determined by two, positive parameters $a$ and $b$.  The size of $d=b^2/a^2$ determines a particularly oriented surface's type, with $d=1$ corresponding to the sphere and psuedosphere, $d>1$ corresponding to the Bulge and Hyperboliod Type, and $d\in(0,1)$ corresponding to the Spindle and Conic Type.

Using (\ref{dilcurve}) and the explicit profile curves, $\gamma(t)$, the dilatation $\mu$ is easily expressed in terms of the parameter $d$ and the height coordinate, $y$, on the circle at infinity.  These expressions are related in elementary ways.  For the Sphere and the Psuedosphere one has \begin{equation} \label{Sphere} \mu_{\text{S}}(y) = \frac{1+y}{3-y} = \mu_{\text{Ps}}(-y) \end{equation} with $y\in[-1,1]$ for the sphere and $y\in[-1,1)$ for the Psuedosphere.  The dilatation for the Bulge and Hyperboloid Type are given by \begin{equation} \label{Bulge} \mu_{\text{B}}(y, d) = \frac{4d-3+2y+y^2}{4d-1+2y-y^2} = \mu_{\text{H}}(-y,d+1) \end{equation} with $d>1$ and $y\in[-1,1]$.  Lastly, the Spindle and the Conic Type have \begin{equation}\label{Spindle} \mu_{\text{Sp}}(y, d) = \frac{4d-3+2y+y^2}{4d-1+2y-y^2} = -\mu_{\text{C}}(-y,1-d) \end{equation} with $d\in(0,1)$ and $y\in[1-2d,1]$ for the Spinde and $y\in[-1,1-2d]$ for the Conic Type.  Note that neither the Spindle nor the Conic type is capable of achieving a full field of view, i.e. image a full hemisphere.

These expressions allow us to easily prove a surprising generalization of the Map-Maker's problem that we call

\begin{thm} \label{GenMapMaker} (The Mirror Maker's Problem) There is no Quasiconformal, Equi-areal Mirror Surface (of revolution). \end{thm}

\proof In each of the expressions above one can determine values of $y$ where $|\mu| = 1$.  These occur at the specified end-points for $y$ and correspond to points where our mirrors either have a vertical tangent plane or a singularity.  For instance, $\mu_{\text{B}}$ and $\mu_{\text{S}}$ have unit length when $y=1$ while $\mu_{\text{Sp}}$ has unit length when $y=1-2d$. $\square$

By restricting the field of view to avoid those points where equations (\ref{Sphere}), (\ref{Bulge}), and (\ref{Spindle}) have unit length, one can limit the amount of angle distortion caused by an equi-areal mirror. These considerations lead us to the following 

\begin{thm} \label{BestFOV} Let $m \in (0,1)$ be fixed.  Of all (portions of) equi-areal mirrors with $$\|\mu\|_{\infty}\leq m$$ the sphere and pseudosphere offer the largest field of view. \end{thm}

\begin{figure}[H]
    \begin{center}
    \includegraphics[width=.75\linewidth]{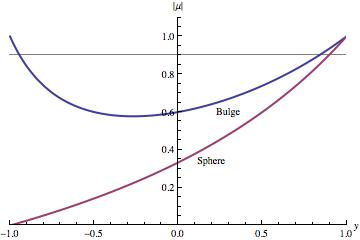}
    \end{center}
    \captionsetup{font=footnotesize}
    \caption[]{Dilatation $|\mu|$ for Sphere and Bulge with $d=1.5$}
    \label{fig:Bulge}
\end{figure}

\proof (sketch) An elementary but careful analysis of the dilatations in (\ref{Sphere}), (\ref{Bulge}), and (\ref{Spindle}) lies at the heart of our proof.  As indicated in Figure \ref{fig:Bulge}, one first observes that to achieve $\|\mu\|_{\infty}=m$, values of $y$ near the end-points are discarded.

The key observation is that as $d\to1$, the surfaces of constant positive (\emph{resp.} negative) Gaussian curvature tend to a sphere (\emph{resp.} psuedosphere), and their fields of view improve as they do so.  

This phenomenon is easy to establish when comparing spheres and bulges (\emph{resp}. psuedospheres and hyperboloid-types).  No matter what value we set for $\|\mu\|_{\infty}=m$, the sphere will always offer a larger field of view than the bulge.  This is depicted in Figure \ref{fig:Bulge} where we have set $\|\mu\|_{\infty}=0.9.$ 

\begin{figure}[H]
    \begin{center}
    \includegraphics[width=.8\linewidth]{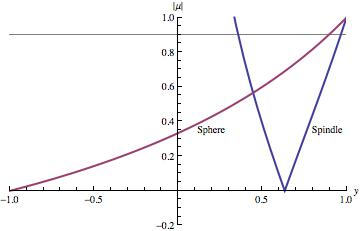}
    \end{center}
    \captionsetup{font=footnotesize}
    \caption[]{Dilatation $|\mu|$ for Sphere and Spindle with $d=1/3$}
    \label{fig:Spindle}
\end{figure}

As indicated in Figure \ref{fig:Spindle}, though, comparing the Sphere and the Spindle (\emph{resp}. the psuedosphere and the conic-type) is more subtle.  The Spindle can offer a larger field of view when $y$ is sufficiently close to $y_f$ where $|\mu(y_f)| = \|\mu\|_{\infty}=m$.  

Solving equation (\ref{Spindle}) for $y$ yields 
$$y = y(\mu_{\text{Sp}}) = \frac {\mu_{\text{Sp}}-1\, + 2\sqrt {1+\left(\mu_{\text{Sp}}\right)^{2}d-d}}{\mu_{\text{Sp}}+1}$$ and from this we can conclude that the largest \emph{total} field of view for the Spindle will occur when we set \begin{align*} y_i &= y(-m) =  \frac {-m-1\, + 2\sqrt {1+{m}^{2}d-d}}{1-m} \\ \\ y_f &= y(m) =  \frac {m-1\, + 2\sqrt{1+{m}^{2}d-d}}{m+1} \end{align*}  Our Spindles will offer a field of view given by $$\theta(d) = \arcsin(y(m)) - \arcsin(y(-m)).$$  A straight-forward computation confirms that $\theta'(d) > 0$, and so the overall field of view increases as $d\to1$.  This means that, given a fixed amount of angle distortion, the Spindle's total field of view is improved as it becomes more spherical.

The relationships between the dilatations for different surfaces in equations (\ref{Sphere}), (\ref{Bulge}), and (\ref{Spindle}) allow us to make analogous conclusions for the negatively curved surfaces.  $\square$

Similar analysis of the dilatations allow us to conclude the converse of Theorem \ref{BestFOV} which we state as a corollary:

\begin{cor} \label{LeastMu} Of all (portions of) equi-areal mirrors that achieve a fixed field of view, $\theta_0$, the sphere and psuedosphere minimize $\|\mu\|_{\infty}$. \end{cor}




\section{Distortion Functionals}\label{Sec:EquitableMirrors}

Given that equi-areal mirrors destroy angles and that parabolic mirrors suffer highly non-uniform magnification, it is natural, then, to determine mirrors that do a fairer or more equitable job of managing both kinds of distortion.  We call such a surface an \emph{equitable mirror}.

Our basic strategy for determining these mirrors involves a two-step process.  First, we devise an appropriate functional, $\mathcal{D}:V\rightarrow\mathbb{R}$, where $V$ is a relevant space of functions.  Second, we find and analyze the critical points of $\mathcal{D}$.

To interpret $\mathcal{D}(v)$ as measuring combined angular and areal distortion, the functional must satisfy basic properties.  The functional should vanish, $\mathcal{D}(v) = 0$, if and only if $v$ is associated to an ideal mirror, one with zero angle and area distortion.  In accordance with Gauss' Theorem Egregium, then, we first and foremost require that $\mathcal{D}(v) > 0$
 for all $v\in V$ whose mirrors offer a non-trivial field of view.  Moreover, formulas for $\mathcal{D}$ should be well-motivated, either incorporating notions of angle distortion, $\mu$, and area distortion, $m_f$, or comparing mirrors to those with desirable properties, like spheres and paraboloids.

In much of what follows, our functionals are defined in terms of $y(r) \in [-1,1]$, the height coordinate on the circle at infinity.  In these cases, potential minimizers are determined by numerically solving pertinent Euler Lagrange equations.  The actual mirror surface is then obtained by determining its profile curve, $f(r)$, from the relation \begin{equation} \label{yf} y = \frac{(f')^2-1}{(f')^2+1}. \end{equation}

When determining the best or least-distortive conic in subsection \ref{BestCon}, our functional is instead phrased in terms of $f(r)$ and is minimized via other, more elementary techniques.  

In any event, to set up a tractable and realistic equitable mirror problem we first assume that the camera's sensor lies along some finite interval.  Without loss of generality we can take this interval to be $[-1,1]$, and, as a result, the spaces $V$ involve sufficiently smooth functions with common domain $[0,1$]. 

\subsection{The Most Equitable Conic} \label{BestCon}
Because they are inexpensive to manufacture and include both the parabaloid and the sphere, conic sections provide us with a natural family of surfaces to explore.  In particular, it is reasonable to suspect that an ellipse or a hyperbola can serve as an equitable mirror.   

Let $V$ denote the space of functions that are continuous on $[0,1]$ and differentiable on $(0,1)$.  Of particular interest are those functions $f(r)\in V$ whose graphs correspond to portions of conic sections.  We assume that our candidate functions satisfy $f'(0)=0$ and $f'(1)=c\in(0,\infty)$ so that a full ($180^\circ$) field of view is obtained if and only if $c=\infty$. 

For this problem we use the rather sensitive functional, $\mathcal{D}_0$, given by \begin{equation}\label{FirstFun} \mathcal{D}_0(f) \coloneqq (K_f-1) + \left\|m_f'\right\|_{\infty} \end{equation} where $K_f$ is the smallest number that satisfies inequality (\ref{quasi}).  We are interested in applying $\mathcal{D}_0$ to graphs of parabolas, hyperbolas, and ellipses and need not consider linear functions since they destroy areas and angles under orthographic projection.  

The $\mathcal{D}_0$-distortion of a hyperbola $$h(r) = \left(c\,a\,\sqrt{1+a^2}\right)\sqrt{1+\frac{r^2}{a^2}}$$ decreases (for any fixed $c \in (0, \infty)$) as $a\to\infty$.  The limiting distortion is easily computed as $$\lim_{a\to\infty}\mathcal{D}_0(h) = 1 + \sqrt{5}\frac{25}{27}c^3 = \mathcal{D}_0(p)$$ which equals the distortion of a parabola, $p(r) = (c/2)r^2$.  This is to be expected since, after appropriate normalizations, our hyperbolae limit on this parabola as $a$ becomes unbounded.  

\begin{figure}[H]
   \begin{center}
   \includegraphics[width=.75\linewidth]{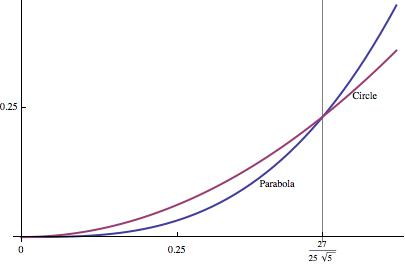}
   \end{center}
   \captionsetup{font=footnotesize}
   \caption[]{$\mathcal{D}_0$-Distortion of Spherical and Parabolic Mirrors}
   \label{fig:SphereParab}
\end{figure}

Similarly, the $\mathcal{D}_0$-distortion of an ellipse $$e(r) = -\left(c\,a\,\sqrt{a^2-1}\right)\sqrt{1-\frac{r^2}{a^2}}$$ decreases as $a\to\sqrt{1+c^{-2}}$, causing our ellipses to become circular.  The limiting distortion coincides with that of a sphere's.  It is given by $$\mathcal{D}_0(s) = c^2$$ which is plotted against the parabola's distortion, $\mathcal{D}_0(p)$ in Figure \ref{fig:SphereParab}.

Figure \ref{fig:SphereParab} illustrates that, for undesirably small fields of view, the parabola offers less $\mathcal{D}_0$-distortion than the circle.  For all $c > 27/(25\sqrt{5})$, a spherical mirror has less $\mathcal{D}_0$-distortion.  With respect to this functional, then, a circle is the best conic section.  

It is natural to regard $\mathcal{D}_0$ as a sort of ``gold-standard" distortion-functional, one that remains finite if and only if mirrors are simultaneously quasiconformal \emph{and} quasi-equi-areal.  However, minimizing $\mathcal{D}_0$ over a wider class of functions is a difficult problem.  Moreover, requiring that our distortion be point-wise bounded is unnecessarily restrictive.  As a result, we focus efforts on minimizing other, equally motivated but more easily managed functionals.

\noindent \subsection{Manageable Distortion} \label{Dist}

The dilatation and magnification factor for a rotationally symmetric mirror can be expressed in terms of $y(r)$ as follows:

\begin{align} \mu &= \frac{ry'-(1+y)(1-y)}{ry'+(1+y)(1-y)} \label{Muy} \\ \nonumber \\ m_f &= \frac{1}{2}\left(\frac{y'(r)}{r}\right) \label{my} \end{align}  Expressions (\ref{Muy}) and (\ref{my}) are obtained by using formulas (\ref{dilgraph}), (\ref{mf}), and (\ref{yf}). 

Note that a mirror will be conformal if the numerator for (\ref{Muy}) is zero, and it will be equi-areal if and only if the derivative of (\ref{my}) also vanishes. This motivates use of the following functional:

\begin{equation} \mathcal{D}_2(y) \coloneqq \int_0^1 \! \left(ry'+y^2-1\right)^2 + \left(ry''-y'\right)^2\,dr \end{equation} where $y\in C^4\left([0,1]\right)$. In order to minimize $\mathcal{D}_2$, we solve the associated Euler-Lagrange equation which is given by

\begin{equation} r^2y^{(4)} + 4r y^{(3)} - r^2 y'' - 2r y'+2y^3 + y^2 - 2y + 1 = 0 \label{EulerLagrange1} \end{equation}   Equation (\ref{EulerLagrange1}) is of fourth order, and so we impose initial conditions, $y(0)=-1, y'(0)=0, y(1)=C$, and $y'(1)=D$.  The first two conditions ensure that $y(r)$ and $f(r)$ are smooth, while the third and fourth conditions respectively control the field of view and magnification factor at $r=1$.  

Although the second variation of $\mathcal{D}_2$ is difficult to work with, we can numerically solve (\ref{EulerLagrange1}) and, likewise, numerically verify that these solutions (locally) minimize $\mathcal{D}_2$; naturally, we carry this out for large values of $C$.  

\begin{figure}[H]
    \begin{center}
    \includegraphics[width=.75\linewidth]{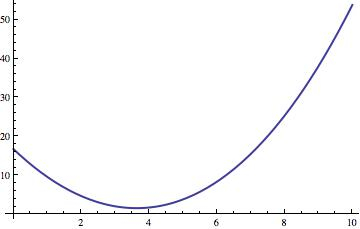}
    \end{center}
    \captionsetup{font=footnotesize}
    \caption[]{$\mathcal{D}_2(y)$ as a function of $D$}
    \label{fig:D2Plot}
\end{figure}

For instance, Figure \ref{fig:D2Plot} depicts the relationship between $\mathcal{D}_2(y)$ and the choice of the magnification parameter, $D$.  In this plot, we have set $y(1)=C=1$ so that our mirrors each achieve a full field of view.  

The minimum occurs at approximately $D\approx 3.636$ where $\mathcal{D}_2(y) \approx 1.60133$.  One readily checks that as $C\to1$, the $\mathcal{D}_2$-distortions of a parabola and hyperbola become unbounded.  The $\mathcal{D}_2$-distortion of an elliptical mirror is strictly greater than that of a spherical one, and, since these findings mirror those for the $\mathcal{D}_0$ functional, this suggests that our $\mathcal{D}_2$ functional is similarly informative.   

A sphere with height coordinate $y_S(r)$ and field of view restricted by $y_S(0)=-1$ and $y_S(1) = C$ has $\mathcal{D}_2$-distortion $$\mathcal{D}_2(y_S) = \frac{(1+C)^4}{9}.$$ When  $C$ is near $1$, then, the sphere's distortion is near $16/9 = 1.77778 > \mathcal{D}_2(y)$.  Hence, for large fields of view, this equitable mirror will feature less $\mathcal{D}_2$-distortion.

\begin{figure}[H]
    \begin{center}
    \includegraphics[width=.75\linewidth]{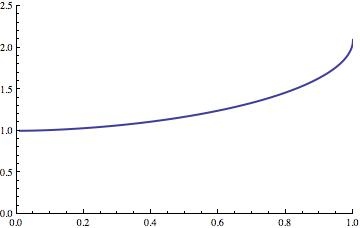}
    \end{center}
    \captionsetup{font=footnotesize}
    \caption[]{$\mathcal{D}_2$ Equitable Mirror (\ref{EulerLagrange1})}
    \label{fig:EquitableMirror1}
\end{figure}

\begin{figure}[H]
    \begin{center}
    \includegraphics[width=.75\linewidth]{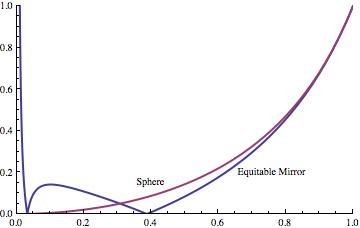}
    \end{center}
    \captionsetup{font=footnotesize}
    \caption[]{$|\mu(r)|$ for the $\mathcal{D}_2$-equitable mirror and Sphere}
    \label{fig:MuSphereEquitable}
\end{figure}

\begin{figure}[H]
    \begin{center}
    \includegraphics[width=.75\linewidth]{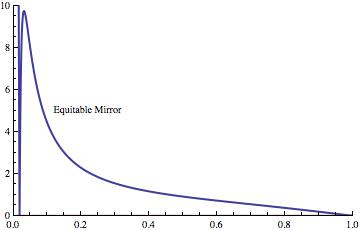}
    \end{center}
    \captionsetup{font=footnotesize}
    \caption[]{$|m_f(r)|$ for the $\mathcal{D}_2$ equitable mirror}
    \label{fig:mfSphereEquitable}
\end{figure}

Figure \ref{fig:EquitableMirror1} shows the profile curve for this mirror, and Figures \ref{fig:MuSphereEquitable} and \ref{fig:mfSphereEquitable} display plots of $|\mu(r)|$ and $|m_f(r)|$, respectively.

Although this equitable mirror attains a smaller amount of distortion than the Sphere, it nonetheless destroys angles (when $r=0$ and when $r=1$).  Furthermore, the magnification factor changes at high rates; in fact, $m_f'$ becomes unbounded near $r=0$. Fortunately, these features occur relatively close to $r=0$ where the catadioptric sensor's camera images itself and therefore has a blind spot.  On average this blind spot occupies approximately 1\% of the image plane, and, as a result, our equitable mirror will feature very little angle distortion and mild area distortion.

\begin{figure}[H]
    \begin{center}
    \includegraphics[width=.75\linewidth]{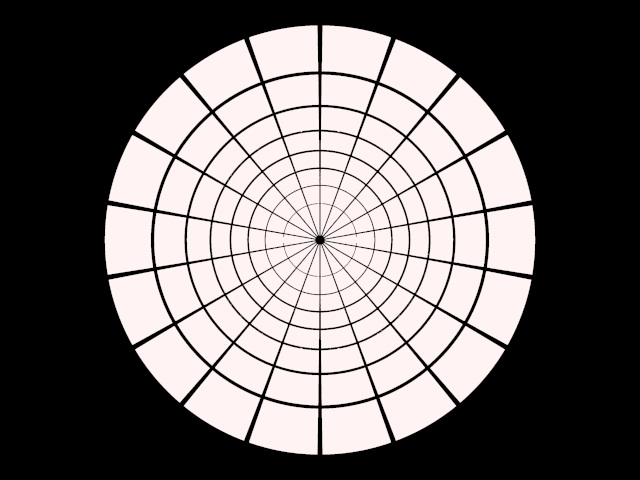}
    \end{center}
    \captionsetup{font=footnotesize}
    \caption[]{Parabolic Mirror in the Angular Test Room}
    \label{fig:parangleroom}
\end{figure}

\begin{figure}[H]
    \begin{center}
    \includegraphics[width=.75\linewidth]{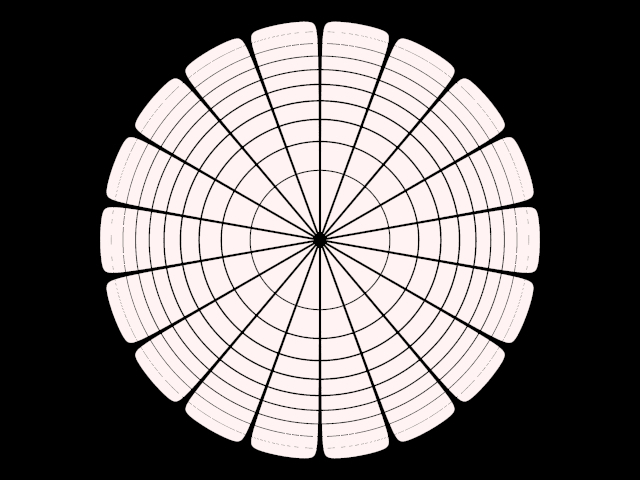}
    \end{center}
    \captionsetup{font=footnotesize}
    \caption[]{Spherical Mirror in the Angular Test Room}
    \label{fig:sphereangleroom}
\end{figure}

To better assess this equitable mirror's strengths and weaknesses, we create three synthetic test rooms using POV-Ray, a ray tracing program.  The first room is spherical with the sensor located at the room's center.  This arrangement allows us to qualitatively assess a mirror's angular distortion by examining how lines of latitude and longitude are imaged.  Figures \ref{fig:parangleroom} and \ref{fig:sphereangleroom}, for instance, show how parabolic and spherical mirrors perform in this room.  

\begin{figure}[H]
    \begin{center}
    \includegraphics[width=.75\linewidth]{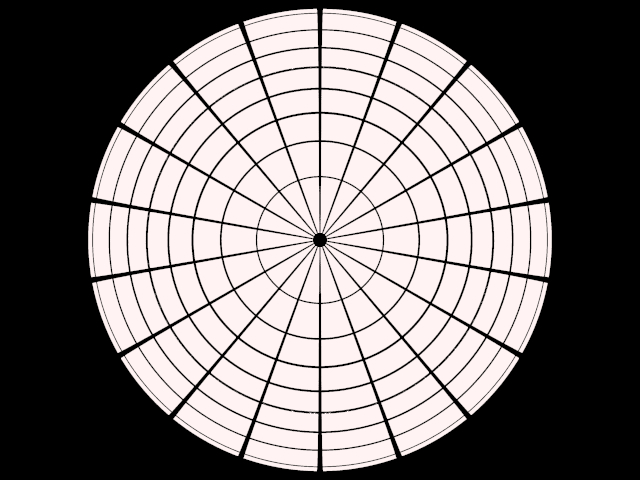}
    \end{center}
    \captionsetup{font=footnotesize}
    \caption[]{The $\mathcal{D}_2$ equitable mirror with in first test room}
    \label{fig:D2intheangulartestroom}
\end{figure}

As expected, the parabolic mirror images latitudinal lines as equally spaced, concentric circles.  This corresponds to the fact that parabolic mirrors preserve angles.  For comparison, Figure \ref{fig:sphereangleroom} shows how a spherical mirror performs in this room.  Note that lines of latitude are imaged as un-equally spaced, concentric circles.  The $\mathcal{D}_2$ equitable mirror has a nearly $270^\circ$ field of view with minimal angle-distortion and reasonable linear distortion as a bonus. Figure \ref{fig:D2intheangulartestroom} depicts how the $\mathcal{D}_2$ mirror performs in the first test room. 

The second test room is rectangular in shape, with walls that are checkered in different colors, and a sensor located in its center. This allows one to qualitatively determine how angles are distorted as well as how lines are imaged by the catadioptric sensor. The third test-room we employ is also rectangular and features seven equal-sized spheres, each equidistant from the catadioptric sensor but of varying elevation. This scene, as used in ~\cite{hp}, allows the observer to determine how the sensor performs in terms of preserving areas. With a true equi-areal sensor, then, the imaged spheres will be distorted in shape but will each have the same area. 

\begin{figure}[H]
    \begin{center}
    \includegraphics[width=.75\linewidth]{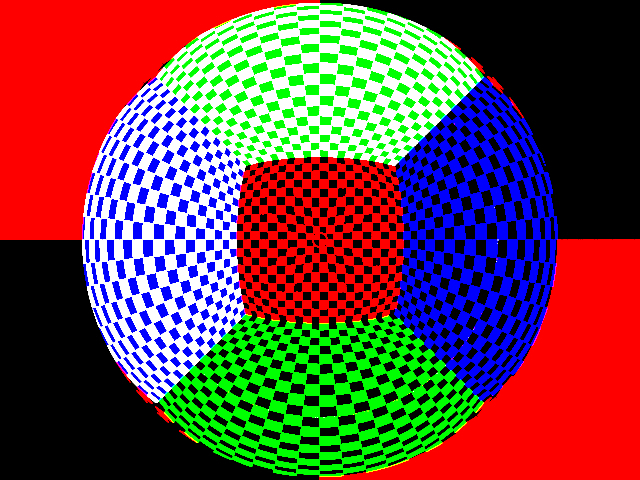}
    \end{center}
    \captionsetup{font=footnotesize}
    \caption[]{The $\mathcal{D}_2$ equitable mirror in second test room.}
    \label{fig:D2-eq-testroom}
\end{figure}

\begin{figure}[H]
    \begin{center}
    \includegraphics[width=.75\linewidth]{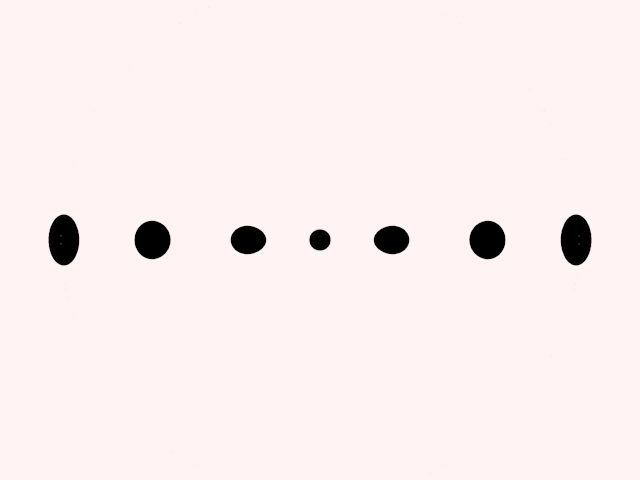}
    \end{center}
    \captionsetup{font=footnotesize}
    \caption[]{The $\mathcal{D}_2$ equitable mirror in third room.}
    \label{fig:D2-eq-sphericaltestroom}
\end{figure}

As is seen in Figure \ref{fig:D2-eq-testroom}, the floor of the second test room (red-black checkered plane) is almost perfectly rescaled. Additionally, Figure \ref{fig:D2-eq-sphericaltestroom} shows that although there is variation in the areas of the imaged spheres, the change is minimal. From these two qualitative analyses, we can conclude that the $\mathcal{D}_2$ equitable mirror is superior to the sphere and the parabolic mirrors.  

Another functional that produces equitable mirrors is defined by \begin{equation}\label{SPFunc} \mathcal{D}_{S,P}(y) \coloneqq \int_0^1 \! \left(y - y_S\right)^2 + \left(y'-y_P'\right)^2\,dr \end{equation}

 where $y_S(r)$ and $y_P(r)$ denote the height coordinates for a sphere and parabola, respectively, with field of view $y(1)=C$. The Euler-Lagrange equation for $\mathcal{D}_{S,P}(y)$ is

\begin{equation}\label{EulerLagrangeSP} y''-y = F(C,r) \end{equation} where $$F(C,r) = 1+(1+C)\left(-r^2+\frac{4((C-1)^2+3(C^2-1)r^2)}{(1+r^2+C(r^2-1))^3}\right).$$  We impose the boundary conditions $y(0)=-1$ and $y(1)=C$, and with extensive use of the Exponential-Integral-Function, explicit solutions to this equation are available.  Given their lengthy and cumbersome appearance, it is best to treat them as numerical solutions.  

\begin{figure}[H]
    \begin{center}
    \includegraphics[width=.75\linewidth]{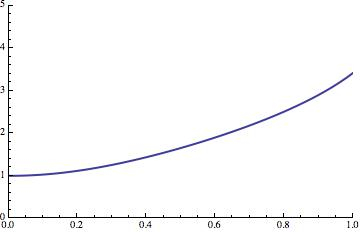}
    \end{center}
    \captionsetup{font=footnotesize}
    \caption[]{Equitable Mirror for $\mathcal{D}_{S,P}$ with $C=.95$}
    \label{fig:SPProfileFullFOV}
\end{figure}

The second variation for $\mathcal{D}_{S,P}$ is easily computed and is always positive.  This means that the critical points we find are local minima.  Figure \ref{fig:SPProfileFullFOV} depicts the profile curve for one such minimum. 

\begin{figure}[H]
    \begin{center}
    \includegraphics[width=.75\linewidth]{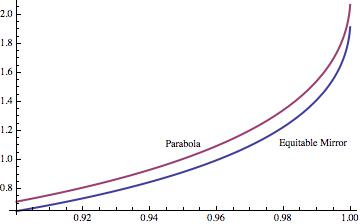}
    \end{center}
    \captionsetup{font=footnotesize}
    \caption[]{$\mathcal{D}_{S,P}$-distortion as a function of $C$}
    \label{fig:DSPParEquit}
\end{figure}

This second equitable mirror outperforms both the sphere and the paraboloid (in terms of $\mathcal{D}_{S,P}$-distortion).  In fact, the $\mathcal{D}_{S,P}$-distortion of a sphere becomes unbounded as $C\to1$, while the distortion values for the paraboloid and the equitable mirror are shown in Figure \ref{fig:DSPParEquit}.

Note that when $C=1$, equation (\ref{EulerLagrangeSP}) simplifies to $$y''- y = 1-2r^2$$ which can be solved explicitly: $$y(r) = 3+2r^2+C_1e^r+C_2e^{-r}.$$  Unfortunately, the $\mathcal{D}_{S,P}$ functional is discontinuous at $C=1$.  The total distortion of the solution above is approximately $4.12108$ which is significantly larger than that of a parabola.

\begin{figure}[H]
    \begin{center}
    \includegraphics[width=.75\linewidth]{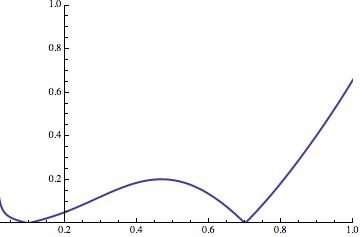}
    \end{center}
    \captionsetup{font=footnotesize}
    \caption[]{$\left|\mu(r)\right|$ for the $\mathcal{D}_{S,P}$ equitable mirror with $C=.95$}
    \label{fig:SPMu}
\end{figure}

\begin{figure}[H]
    \begin{center}
    \includegraphics[width=.75\linewidth]{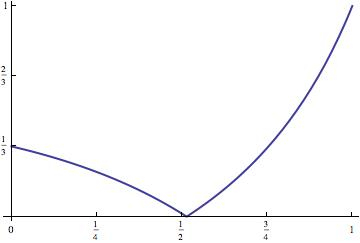}
    \end{center}
    \captionsetup{font=footnotesize}
    \caption[]{$\left|\mu(r)\right|$ for the $\mathcal{D}_{S,P}$ equitable mirror with $C=.95$}
    \label{fig:SPmuFullFOV}
\end{figure}

Moreover, as shown in Figures \ref{fig:SPMu} and \ref{fig:SPmuFullFOV}, the complex dilatation remains bounded above by 1 (and below by -1), except when $C=1$ in which case $\mu(1) = 1$.  

The magnification factor behaves similarly to the magnification factor for the $\mathcal{D}_2$ equitable mirror when $C<1$.  


\begin{figure}[H]
    \begin{center}
    \includegraphics[width=.75\linewidth]{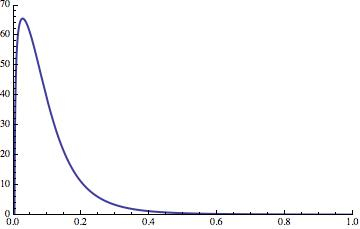}
    \end{center}
    \captionsetup{font=footnotesize}
    \caption[]{$\left|mf(r)\right|$ for the $\mathcal{D}_{S,P}$ equitable mirror with $C=.95$}
    \label{fig:SPmfPrime}
\end{figure}

\begin{figure}[H]
    \begin{center}
    \includegraphics[width=.75\linewidth]{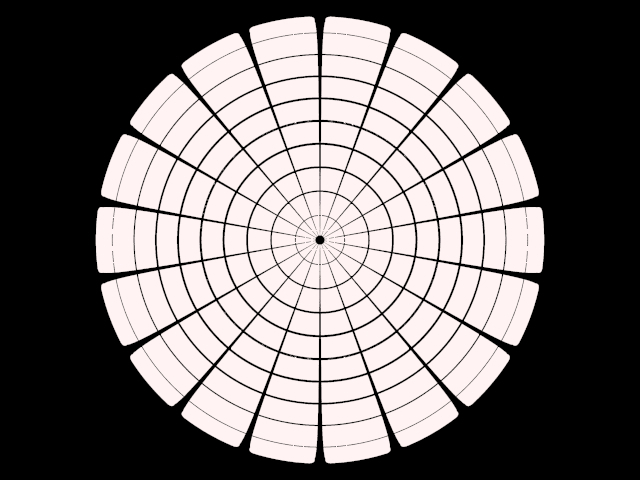}
    \end{center}
    \captionsetup{font=footnotesize}
    \caption[]{$\mathcal{D}_{SP}$ equitable mirror with $C= .95$ in first test room.}
    \label{fig:DSPAngleRoom}
\end{figure}
\begin{figure}[H]
    \begin{center}
    \includegraphics[width=.75\linewidth]{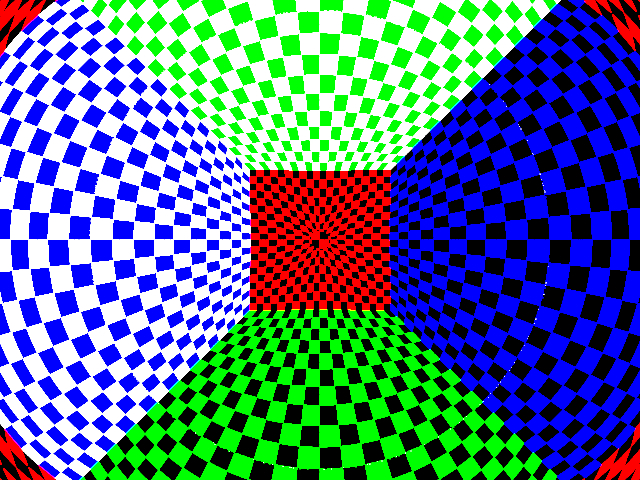}
    \end{center}
    \captionsetup{font=footnotesize}
    \caption[]{$\mathcal{D}_{SP}$ equitable mirror with $C= .95$ in second test room.}
    \label{fig:DSP-eq-testroom}
\end{figure}
\begin{figure}[H]
    \begin{center}
    \includegraphics[width=.75\linewidth]{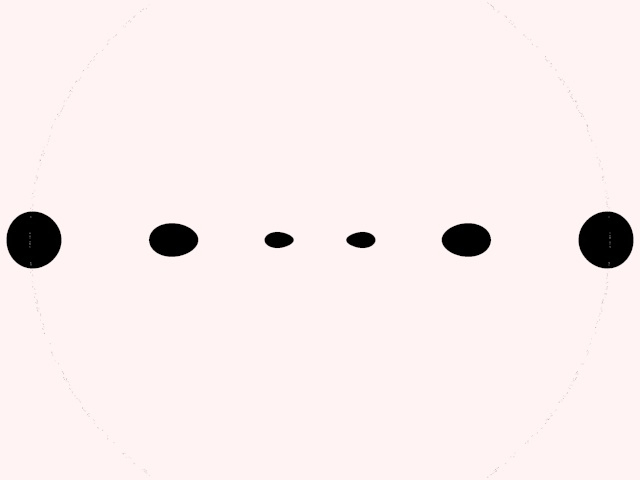}
    \end{center}
    \captionsetup{font=footnotesize}
    \caption[]{$\mathcal{D}_{SP}$ equitable mirror with $C= .95$ in third test room.}
    \label{fig:DSP-eq-testroom-spherical}
\end{figure}

The performance of the $\mathcal{D}_{S,P}$ equitable mirror can be seen in Figures \ref{fig:DSPAngleRoom}, \ref{fig:DSP-eq-testroom} and \ref{fig:DSP-eq-testroom-spherical}. According to the second and third test rooms, it is inferior to the $\mathcal{D}_2$ equitable mirror, even though the angles are reasonably preserved.  In particular, areas are severely distorted by this mirror.

It should be noted that the analogously-defined functional $$\mathcal{D}_{P,S}(y) \coloneqq \int_0^1 \! \left(y_p-y\right)^2+\left(y_S'- y'\right)^2\,dr$$ can be (locally) minimized via identical methods.  However, the minima fail to feature less distortion than a spherical mirror.

\section{Summary of Results and Future Work}
We summarize our main results as follows.

\begin{enumerate}
\item Spherical and Psuedo-spherical mirrors offer the widest fields of view amongst all equi-areal mirrors with restricted angle distortion (as established in Theorem \ref{GenMapMaker}).
\item Spherical and Psuedo-spherical mirrors offer the least amount of angle distortion amongst all equi-areal mirrors with a specified field of view (as established in Corollary \ref{LeastMu}).
\item The unexplored problem of simultaneously minimizing multiple kinds of distortion can be elegantly posed and solved via distortion functionals and standard calculus-of-variations techniques.
\item The equitable mirror obtained by (numerically) minimizing the $\mathcal{D}_2$ functional is one solution to the problem of simultaneously managing angle and area distortion.  This new mirror performs quite well in various test rooms. 
\end{enumerate}

There are many directions in which to continue this research.  Perhaps the most obvious concerns the development of other distortion functoinals; different applications and problems will motivate different distortion and cost functionals.  For instance, one can easily generalize the definition of $\mathcal{D}_2$ to the more general \begin{equation} \label{DPFunc} \mathcal{D}_p(y) \coloneqq \int_0^1 \! \left|ry'+y^2-1\right|^p + \left|ry''-y'\right|^p\,dr\end{equation} where $p\geq 1$.  The ensuing Euler-Lagrange equations are significantly difficult, but for the sphere one readily computes $$\mathcal{D}_p(y_S) = \frac{(1+C)^{2p}}{(1+4p)}$$  Since the $\mathcal{D}_p$ functionals become more sensitive as $p\to\infty$, solutions to the associated Euler Lagrange equations may provide equitable mirrors with even more desirable properties. 

It is possible that in order to solve certain problems both angle and area distortion need to be simultaneously minimized, but one kind of distortion is more important or highly prioritized than the other.  In such a scenario, the functional $$\int_0^1 \! \left(ry'+y^2-1\right)^{2\alpha}+\left(ry''-y'\right)^2\,dr$$ might be of service since it is more sensitive to angle distortion (for $\alpha > 1$) than it is to area distortion.

Alternatively, one may instead want to minimize different kinds of distortion altogether, including (but not limited to) distortions in distances and resolution.  Additionally, the actual cost of producing a mirror can be factored in to functionals.  For example, if we assume that smaller mirrors are less expensive to produce, then an expression like $$\int_0^1 \! \left(ry'+y^2-1\right)^2+\left(ry''-y'\right)^2 + \sqrt{\frac{2}{1-y}}\,dr$$ would be of use since it also takes into account the arc-length of the mirror's profile curve.


\section*{Acknowledgements}
This work was partially funded by the Center for Undergraduate Research (CURM) and NSF grant \#DMS-1148695.
\ \\

\end{document}